\documentclass[prd,showpacs,amsmath,amssymb]{revtex4}
\usepackage{dcolumn}% Align table columns on decimal point
\usepackage{bm}% bold math
\usepackage{multirow}
\usepackage{epsfig,graphicx}
\begin{document}
\title{Open charm effects in the explanation of the long-standing ``$\rho\pi$ puzzle" }
\author{Qian Wang$^1$, Gang Li$^3$ and Qiang Zhao$^{1,2}$}

\affiliation{1) Institute of High Energy Physics, Chinese Academy of
Sciences, Beijing 100049, P.R. China \\
2) Theoretical Physics Center for Science Facilities, CAS, Beijing
100049, P.R. China \\
3) Department of Physics, Qufu Normal University, Qufu, 273165, P.R.
China}

\date{\today}

\begin{abstract}
A detailed analysis of the open charm effects on the decays of
$J/\psi(\psi^\prime)\to VP$ is presented, where $V$ stands for light
vector meson and $P$ for light pseudoscalar meson. These are the
channels that the so-called ``12\% rule" of perturbative QCD (pQCD)
is obviously violated. Nevertheless, they are also the channels that
violate the pQCD helicity selection rule (HSR) at leading order. In
this work, we put constraints on the electromagnetic (EM)
contribution, short-distance contribution from the $c\bar{c}$
annihilation at the wavefunction origin, and long-distance
contribution from the open charm threshold effects on these two
decays. We show that interferences among these amplitudes, in
particular, the destructive interferences between the short-distance
and long-distance strong amplitudes play a key role to evade the HSR
and cause the significant deviations from the pQCD expected ``12\%
rule".

\end{abstract}

\date{\today}

%13.25.Gv Decays of J/¦×, ¦´, and other quarkonia

%12.38.Lg Other nonperturbative calculations

%12.40.Vv Vector-meson dominance

\maketitle

\section{Introduction}

Annihilation decays of heavy quarkonium have served as an important
probe for the study of the perturbative QCD (pQCD) strong
interactions in the literature
\cite{Duncan:1980qd,Brodsky:1981kj,Chernyak:1981zz}. In the
bottomonium energy region, the non-relativistic approximation works
well so that the annihilation of the $b\bar{b}$ can be regarded as a
direct measurement of the properties of the bottomonium
wavefunctions at the origin at leading order. For instance, for the
$S$ wave states, the annihilation matrix elements are proportional
to the wavefunction at the origin, while for the $P$ wave states to
the first derivative at the origin. These simple relations have been
broadly examined and found in good agreement with the experimental
measurements in inclusive processes. They can be regarded as a
direct test of the pQCD properties. Interestingly, although the mass
of the charm quark cannot be regarded heavy enough, some of the
leading pQCD relations are still well respected in inclusive
transitions. A good example is the branching ratio fraction between
$\psi^\prime$ and $J/\psi$:
\begin{eqnarray}
R\equiv\frac{ BR(\psi^\prime\to hadrons)}{BR(J/\psi \to
hadrons)}\simeq\frac{ BR(\psi^\prime\to e^+e^-)}{BR(J/\psi \to
e^+e^-)}\simeq 0.13 \ ,
\end{eqnarray}
which is the so-called ``12\% rule" and the branching ratio
fractions probe the ratio of the wavefunctions at their origins for
the ground state $J/\psi$ and first radial excitation $\psi^\prime$.
Note that in the above equation both branching ratios $BR(J/\psi \to
hadrons)$ and $BR(\psi^\prime\to hadrons)$ are referred to their
light hadron decays. In fact, even for some of those exclusive
decays, the above relation seems to hold approximately well. Such an
observation, in contrast with the significant deviations in $J/\psi$
and $\psi^\prime\to \rho\pi$, has initiated tremendous interests in
the study of transition mechanisms for $J/\psi$ and $\psi^\prime\to
\rho\pi$, which is known as the so-called ``$\rho\pi$ puzzle".
According to the Particle Data Group 2010~\cite{Nakamura:2010zzi},
the ratio for the $\rho\pi$ channel is
$BR(\psi^\prime\to\rho\pi)/BR(J/\psi\to \rho\pi)\simeq (1.1\sim
2.8)\times 10^{-3}$, which is much smaller than the pQCD expected
value, i.e. $\sim 12\%$.

An alternative expression for the ``$\rho\pi$ puzzle" is related to
the power law suppression due to the pQCD helicity selection rule
(HSR). As demonstrated in
Refs.~\cite{Brodsky:1981kj,Chernyak:1981zz}, the decay of $J/\psi \
(\psi')\to VP$, where $V$ and $P$ stand for vector and pseudoscalar
meson, respectively, should be strongly suppressed at leading twist.
As a consequence, the branching ratio fraction is expected to be
$BR(\psi^\prime\to\rho\pi)/BR(J/\psi\to
\rho\pi)\simeq(M_{J/\psi}/M_{\psi^\prime})^6\times (BR(\psi'\to
hadrons)/BR(J/\psi\to hadrons))\sim 6.2\%$, which is still much
larger than the experimental observations. The significant violation
of the pQCD HSR is nontrivial taking into account that quite many
exclusive decay channels have approximately respected the $12\%$
rule.

Such a conflicting phenomenon has attracted a lot of attention from
both experiment and theory in history. Even right now, the study of
the ``$\rho\pi$ puzzle" has been one of the most important physics
goals in the program of BESIII experiment~\cite{bes-iii}. In theory,
this puzzle has also been  broadly studied. Different explanations
have been proposed in the literature, such as the color-octet
model~\cite{Chen:1998ma}, vector meson
mixing~\cite{Clavelli:1983rk,Pinsky:1989ue}, final state
interactions~\cite{Li:1996yn,Suzuki:1998ea}, admixtures of a vector
glueball near $J/\psi$~\cite{Chan:1999px,Gu:1999uq}, intrinsic charm
in light mesons~\cite{Brodsky:1997fj}, light-quark mixing
effects~\cite{Feldmann:2000hs}, large survival decay of $\psi'$ via
virtual charmonium state~\cite{Gerard:1999uf}, and interferences
between the electromagnetic (EM) and strong
interactions~\cite{Li:2007ky,Zhao:2006gw,Seiden:1988rr,Suzuki:2001fs}.
In the meantime, it has been realized that the ``$\rho\pi$ puzzle"
is not just restricted to the $\rho\pi$ decay channel. It has also
connections with the obvious charge asymmetries observed in
$\psi^\prime\to K^*\bar{K}+c.c.$ Therefore, it was conjectured that
more general dynamic reasons should be investigated for $J/\psi \
(\psi^\prime)\to
VP$~\cite{Li:2007ky,Zhao:2008eg,Zhao:2010ja,Mo:2006cy,Li:2008ey}.

It should be useful to recall the results of Ref.~\cite{Li:2007ky},
where a global fit for $J/\psi(\psi^\prime)\to VP$ is presented. The
EM and strong transition amplitudes are parameterized out for all
the decay channels, while among the strong transition amplitudes,
the singly disconnected OZI (SOZI) processes and doubly disconnected
OZI (DOZI) processes are further parameterized out. As shown in
Ref.~\cite{Li:2007ky}, there exists an overall suppression on the
strong decay amplitudes of $\psi^\prime\to VP$, not just in the
$\rho\pi$ channel. Due to this suppression, the EM transition
amplitudes become compatible with the strong decay amplitudes with
which the interferences produce further deviations from the
HSR-violating power law suppressions. This fitting result at least
clarifies the following two issues: i) The same mechanism that
suppresses $\psi^\prime\to \rho\pi$ also plays a role in other
$\psi^\prime\to VP$ decays; ii) Such a mechanism does not affect
much in $J/\psi\to VP$ as suggested by the charge asymmetries
observed in $K^*\bar{K}+c.c.$  These are important guidance for
exploring mechanisms that would suppress the strong decay amplitudes
in $\psi^\prime\to VP$, but have less impact on the $J/\psi$ decays.

During the past few years, we have been focussing on the study of
mechanisms evading the HSR in charmonium decays. For charmonia below
the open $D\bar{D}$ threshold, the HSR violating transitions are
naturally correlated with the OZI-rule violations. As demonstrated
in a series of
studies~\cite{Liu:2009vv,Liu:2010um,Wang:2010iq,Wang:2012wj,Zhang:2009kr}
, we have shown that the intermediate $D$ meson loops (IML) provide
a natural mechanism for evading the OZI rule and hence the HSR in
charmonium decays. The IML is introduced as a non-perturbative
source of contributions. As iterated in Refs.
\cite{Zhao:2008eg,Zhao:2010ja,Zhang:2009kr,Wang:2010iq,Li:2011ss},
apart from the ``$\rho\pi$ puzzle" the IML could be a key for
understanding some of those long-standing questions in charmonium
exclusive decays, e.g. the $\psi(3770)$ non-$D\bar{D}$ decay, large
HSR-violating decay of $\eta_c\to VV$, M1 transition problems with
$J/\psi \ (\psi^\prime)\to \gamma \eta_c (\eta_c^\prime)$, etc.

In this work, we provide a quantitative study of the role played by
the long-distance IML in $J/\psi \ (\psi^\prime)\to VP$ in
association with the EM and short-distance SOZI transitions. Our
purpose is to demonstrate that the IMLs as a non-perturbative
transition mechanism are important for explaining the phenomena
observed in $J/\psi \ (\psi^\prime)\to VP$, hence could be a natural
solution for the long-standing ``$\rho\pi$ puzzle" and other puzzles
in charmonium exclusive decays. As follows, the details of dealing
with different transition amplitudes are given in Sec. II. The
numerical results and detailed analysis are presented in Sec. III,
and a summary in the last section.

\section{The model}

A unique feature with the $VVP$ coupling is that at hadronic level
the anti-symmetric tensor coupling is the only allowed Lorentz
structure. Therefore, it can be understood that whatever the
underlying mechanisms could be, they will contribute to the
corrections to the anti-symmetric tensor coupling. Based on this,
one can always make a general parametrization to the transition
amplitude,
\begin{eqnarray}\label{total-trans}
\mathcal{M}_{tot}\equiv
\mathcal{M}_{EM}+e^{i\delta_0}(\mathcal{M}_{short}+e^{i\theta}\mathcal{M}_{long}
) ,
\end{eqnarray}
where $\mathcal{M}_{EM}$, $\mathcal{M}_{short}$ and
$\mathcal{M}_{long}$ are the amplitudes of the EM, strong
short-distance and strong long-distance transitions. A phase angle
$\theta$ is introduced between the short and long-distance
amplitudes, while the relative phase between the EM and
short-distance amplitudes is $\delta_0=0^\circ$ or $180^\circ$. It
is reasonable to consider the trivial relative phase angles between
the EM and short-distance amplitude. Meanwhile, the long-distance
amplitude may carry a phase angle relative to the short-distance one
due to hadronic wavefunction effects. Although the exclusive
amplitudes for these three sources are obtained as real numbers, the
relative phase angle $\theta$ can lead to a complex coupling in
$J/\psi (\psi')\to VP$. We note that the EM amplitudes for each
decay modes carry intrinsic signs deduced in the quark
model~\cite{Seiden:1988rr}. Our efforts as follows are to constrain
these amplitudes  and present an overall prescription for $J/\psi
(\psi') \to VP$.

\subsection{EM transition amplitudes}

The EM transition $J/\psi (\psi') \to \gamma^* \to VP$ turns out to
be important in $J/\psi (\psi') \to  VP$. In particular, it is the
dominant contribution to those isospin-violating decay channels,
i.e. $J/\psi (\psi^\prime)\to \rho\eta$, $\rho\eta^\prime$,
$\omega\pi^0$ and $\phi\pi^0$. This mechanism can be investigated in
the vector meson dominance (VMD) model as presented in
Refs.~\cite{Zhao:2006gw,Li:2007ky}.

\begin{figure}
\begin{center}
\vglue-0mm
\includegraphics[width=1.0\textwidth]{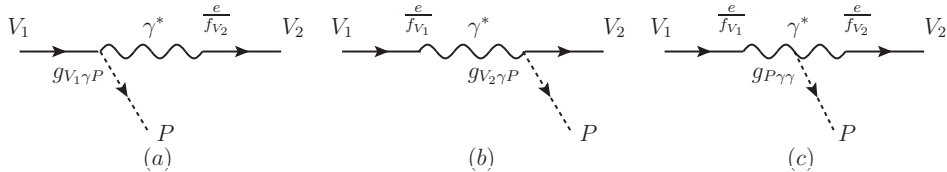}
\vglue-0mm \vspace{-17cm} \caption{The tree-level Feynman diagrams
of EM transitions in $J/\psi(\psi^\prime)\to VP$. }\label{fig-1}
\end{center}
\end{figure}

In Fig.~\ref{fig-1} those three independent electromagnetic
transition processes in the VMD are illustrated. The vertex
couplings can be extracted  from the experimental data for the decay
widths of $V\to\gamma P$ (or $P\to\gamma V$), and $P\to
\gamma\gamma$. However, since the intermediate photon is off-shell,
a form factor $\mathcal{F}(q^2)=\Lambda_{EM}^2/(\Lambda_{EM}^2-q^2)$
is adopted for the EM transition amplitudes. The cut-off energy
$\Lambda_{EM}$ is universal for both $J/\psi$ and $\psi^\prime$
decays, and to be determined by experimental data for those
isospin-violating decay channels. The EM amplitude can thus be
expressed as
\begin{eqnarray}
\nonumber
\mathcal{M}_{EM}&=&\mathcal{M}_a+\mathcal{M}_b+\mathcal{M}_c\\
&=&\left(\frac{e}{f_{V_2}}\frac{g_{V_1\gamma
P}}{M_{V_1}}\mathcal{F}_a+\frac{e}{f_{V_1}}\frac{g_{V_2\gamma
P}}{M_{V_2}}\mathcal{F}_b+\frac{e^2}{f_{V_1}f_{V_2}}\frac{g_{P\gamma\gamma
}}{M_{P}}\mathcal{F}_c\right)\epsilon_{\mu\nu\alpha\beta}p^\mu\epsilon(p)^\nu
k^\alpha\epsilon(k)^\beta,
\end{eqnarray}
where $p$($k$) is the four momentum of the initial vector charmonium
(final light vector), and $\epsilon(p)$ ($\epsilon(k)$) is its
corresponding polarization vector. In Tables~\ref{tab-vgp},
\ref{tab-pgg} and  \ref{tab-vg} the EM vertex couplings are
extracted with the up-to-date data from the
PDG2010~\cite{Nakamura:2010zzi}.

\begin{table}
\caption{The couplings $g_{V\gamma P}\equiv [{12\pi
M_V^2\Gamma(V\to\gamma P)}/{|\mathbf{p}_\gamma|^3}]^{1/2}$ or $
g_{V\gamma P}\equiv [{4\pi M_V^2\Gamma(P\to\gamma
V)}/{|\mathbf{p}_\gamma|^3}]^{1/2}$ determined by experimental data
from PDG2010~\cite{Nakamura:2010zzi}.}
\begin{tabular}{ccc}
  \hline\hline
  $g_{V\gamma P}$ & Values & Branching ratios \\
  \hline
  $g_{\rho\gamma\eta}$ & 0.381 & $(3.00\pm0.20)\times10^{-4}$ \\
  \hline
  $g_{\rho\gamma\eta^\prime}$ & 0.295 & $(29.3\pm 0.5)\%$ \\
  \hline
  $g_{\rho^0\gamma\pi^0}$ & 0.196 & $(6.0\pm0.8)\times10^{-4}$ \\
  \hline
  $g_{\rho^\pm\gamma\pi^\pm}$ & 0.170 & $(4.5\pm 0.5)\times10^{-4}$ \\
  \hline
  $g_{\omega\gamma\eta}$ & 0.107 & $(4.6\pm0.4)\times10^{-4}$ \\
  \hline
  $g_{\omega\gamma\eta^\prime}$ & 0.101 & $(2.75\pm0.22)\%$ \\
  \hline
  $g_{\omega\gamma\pi}$ &0.545 & $(8.28\pm0.28)\%$ \\
  \hline
  $g_{\phi\gamma\eta}$ & 0.214 & $(1.309\pm0.024)\%$ \\
  \hline
  $g_{\phi\gamma\eta^\prime}$ & 0.221 & $(6.25\pm0.21)\times 10^{-5}$ \\
  \hline
  $g_{\phi\gamma\pi}$ & 0.041 & $(1.27\pm0.06)\times10^{-3}$ \\
  \hline
  $g_{K^{*\pm}\gamma K^\pm}$ & 0.226 & $(9.9\pm0.9)\times10^{-4}$ \\
  \hline
  $g_{K^{*0}\gamma \bar{K}^0}$ & 0.344 & $(2.39\pm0.21)\times10^{-3}$ \\
  \hline
  $g_{J/\psi\gamma\eta}$ & $3.31\times10^{-3}$ & $(1.104\pm0.034)\times10^{-3}$ \\
  \hline
  $g_{J/\psi\gamma\eta^\prime}$ & $8.04\times10^{-3}$ & $(5.28\pm0.15)\times10^{-3}$ \\
  \hline
  $g_{J/\psi\gamma\pi}$ & $5.64\times10^{-4}$ & $(3.49^{+0.33}_{-0.30})\times10^{-5}$ \\
  \hline
  $g_{\psi^\prime\gamma\eta}$ & $2.31\times10^{-4}$ & $<2\times10^{-6}$ \\
  \hline
  $g_{\psi^\prime\gamma\eta^\prime}$ & $1.93\times10^{-3}$ & $(1.21\pm0.08)\times10^{-4}$ \\
  \hline
  $g_{\psi^\prime\gamma\pi^0}$ & $3.534\times 10^{-4}$ & $<5.0\times
  10^{-6}$\\
  \hline\hline
\end{tabular}
\label{tab-vgp}
\end{table}

\begin{table}
\caption{The couplings $g_{P\gamma\gamma}\equiv ({32\pi
\Gamma(P\to\gamma\gamma)}/{M_P})^{1/2}$ determined by experimental
data from PDG2010~\cite{Nakamura:2010zzi}. }
\begin{tabular}{cccc}
  \hline\hline
  $g_{P\gamma\gamma}$ & Values &$\Gamma_{tot}(\mathrm{keV})$& Branching ratios  \\
  \hline
  $g_{\pi\gamma\gamma}$ & $2.40\times10^{-3}$ &$7.86\times 10^{-3}$& $(98.823\pm0.034)\%$ \\
  \hline
  $g_{\eta\gamma\gamma}$ & $9.68\times10^{-3}$ &1.3& $(39.31\pm0.20)\%$ \\
  \hline
  $g_{\eta^\prime\gamma\gamma}$ & $2.13\times10^{-2}$ &194& $(2.22\pm0.08)\%$ \\
  \hline\hline
\end{tabular}
\label{tab-pgg}
\end{table}

\begin{table}
\caption{The couplings ${e}/{f_V}\equiv [{3\Gamma_{V\to e^+
e^-}}/{(2\alpha_e|\mathbf{p}_e|)}]^{1/2}$ determined by experimental
data from PDG2010~\cite{Nakamura:2010zzi}. }
\begin{tabular}{cccc}
  \hline\hline
  $e/f_V$ & Values($\times10^{-2}$) & $\Gamma_{tot}(\mathrm{MeV})$ & $BR(V\to e^+e^-)$ \\
  \hline
  $e/f_\rho$ & 6.11 & 149.1 & $(4.72\pm0.05)\times10^{-5}$ \\
  \hline
  $e/f_\omega$ & 1.80 & 8.49 & $(7.28\pm0.14)\times10^{-5}$ \\
  \hline
  $e/f_\phi$ & 2.25 & 4.26 & $(2.954\pm0.03)\times10^{-4}$ \\
  \hline
  $e/f_{J/\psi}$ & 2.71 & 0.0929 & $(5.94\pm0.06)\%$ \\
  \hline
  $e/f_{\psi^\prime}$ & 1.62 & 0.304 & $(7.72\pm0.17)\times10^{-3}$ \\
  \hline\hline
\end{tabular}
\label{tab-vg}
\end{table}

\subsection{Short-distance transition amplitudes}

\begin{figure}
\begin{center}
\hspace{-2cm}
\includegraphics[scale=0.7]{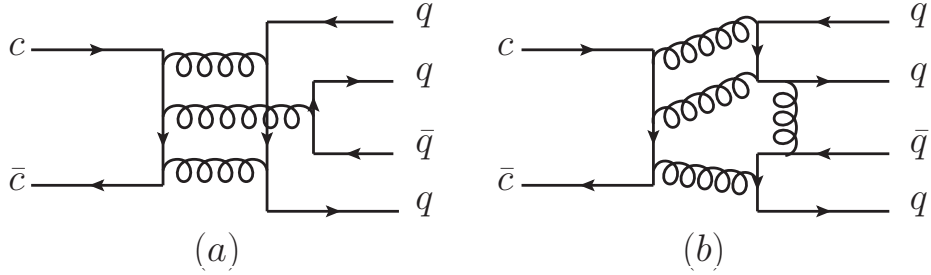}
\vspace{-14cm} \caption{Schematic diagrams for the short-distance
strong transitions in $J/\psi \ (\psi')\to VP$. Diagram (a)
illustrates the SOZI process, while (b) is for the DOZI one. In both
cases,  $c$ and $\bar{c}$ annihilate at the origin of the
wavefunction. }
 \label{short}
\end{center}
\end{figure}

The short-distance contribution of strong interaction is mainly from
the $c\bar{c}$ annihilation at the wavefunction origin associated
with hard gluon radiations. This is an SOZI transition and can be
parameterized out in a similar way as in Ref.~\cite{Li:2007ky}. We
emphasize the $c\bar{c}$ annihilation at the wavefunction origin in
this process. Thus, the HSR violation can be regarded as being
produced by the non-negligible light quark masses in the
hadronization process. The inclusive gluon annihilation part is thus
guaranteed to scale with their lepton pair branching ratio fraction
since the inclusive amplitudes would be controlled by the quarkonium
wavefunction at the origin~\cite{Brodsky:1981kj}. This process
distinguishes from the long-distance transitions via the IML where
the $c\bar{c}$ annihilations would occur non-locally and probe the
charmonium wavefunction away from the origin. Such a difference
would allow us to treat the short and long-distance amplitudes
individually and to avoid double-counting between these two
mechanisms. A schematic diagram for the short-distance SOZI
transitions is shown in Fig.~\ref{short}(a).

The parametrization of the short-distance amplitudes is outlined as
follows~\cite{Li:2007ky}. First, the strength of the non-strange
SOZI process is parameterized as
\begin{eqnarray}
g_{J/\psi(\psi')} =\langle (q\bar q)_V (q\bar q)_P |V_0|
J/\psi(\psi')\rangle,
\end{eqnarray}
where $V_0$ is the $3g$ decay potential of the charmonia into two
non-strange $q\bar q$ pairs of vectors and pseudoscalars via SOZI
processes. But it should be noted that the subscript $V$ and $P$
here do not mean that the quark-antiquark pairs are the SU(3) flavor
eigenstates of vector and pseudoscalar mesons. The amplitude
$g_{J/\psi (\psi')}$ is proportional to the charmonium wavefunctions
at origin. Thus, it may have different values for $J/\psi$ and
$\psi'$.

Considering the SU(3) flavor symmetry breaking, which distinguishes
the $s$ quark pair production from the $u$, $d$ quarks in the
hadronizations, we introduce the SU(3) flavor symmetry breaking
parameter $\xi$,
\begin{eqnarray}
\xi \equiv  \langle
(q\bar{s})_V(s\bar{q})_P|V_0|J/\psi(\psi^\prime)\rangle /
g_{J/\psi(\psi^\prime)} = \langle
(s\bar{q})_V(q\bar{s})_P|V_0|J/\psi(\psi^\prime)\rangle /
g_{J/\psi(\psi^\prime)}
\end{eqnarray}
where $\xi=1$ is in the SU(3) flavour symmetry limit, while
deviations from unity implies the SU(3) flavor symmetry breaking. In
general, the value of parameter $\xi$ is around $\xi\simeq
f_\pi/f_K=0.838$, which provides a guidance for the SU(3) flavor
symmetry breaking effects. For the production of two $s\bar{s}$
pairs via the SOZI potential, the recognition of the SU(3) flavor
symmetry breaking in the transition is accordingly
\begin{eqnarray}
\xi^2 = \langle
(s\bar{s})_V(s\bar{s})_P|V_0|J/\psi(\psi^\prime)\rangle /
g_{J/\psi(\psi^\prime)} \ .
\end{eqnarray}

For the $J/\psi$ and $\psi'$ decays into isoscalar final states,
such as $\omega\eta$, $\omega\eta'$, $\phi\eta$ and $\phi\eta'$, the
DOZI transition as illustrated by Fig.~\ref{short}(b) may also
contribute. Although it is not apparent that the DOZI transition can
be classified as a short-distance process, we can parameterized it
out as follows,
\begin{equation}\label{eq_dozi}
r\equiv \langle
(s\bar{s})_V(q\bar{q})_P|V_0|J/\psi(\psi^\prime)\rangle /
g_{J/\psi(\psi^\prime)} = \langle
(q\bar{q})_V(s\bar{s})_P|V_0|J/\psi(\psi^\prime)\rangle /
g_{J/\psi(\psi^\prime)} \ ,
\end{equation}
of which a small value $|r|<<1$ would suggest a short-distance
nature of this process. We mention that the DOZI process
topologically does not double-count the long-distance IML
transitions to be defined in the next Subsection.

To take into account the size effects of the initial and final state
mesons, a commonly adopted form factor is included, i.e.
\begin{eqnarray}
{\cal {F}}({\bf P}) \equiv |{\bf P}|^{l}\exp ({-{\bf
P}^2/{16\beta^2}})
\end{eqnarray}
where $|{\bf P}|$ is the three-vector momentum of the final-state
mesons in the $J/\psi(\psi^\prime)$ rest frame, and $l$ is the
final-state relative orbital angular momentum quantum number. We
adopt $\beta = 0.5 \ \mbox {GeV}$, which is the same as
Refs.~\cite{Amsler:1995td,Close:2000yk,Close:2005vf}. At leading
order the decays of $J/\psi(\psi^\prime)\to VP$ are via $P$-wave,
i.e. $l=1$.

The transition amplitudes for $J/\psi(\psi^\prime)\to VP$ via the
short-distance SOZI transitions can then be expressed as
\begin{eqnarray}
\mathcal{M}_S(\rho^0\pi^0)&=&
\mathcal{M}_S(\rho^+\pi^-)=\mathcal{M}_S(\rho^-\pi^+)=g_{J/\psi(\psi^\prime)}{\cal
{F}}({\bf P}) \nonumber \\
\mathcal{M}_S(K^{*+}K^-)&=&
\mathcal{M}_S(K^{*-}K^+)=\mathcal{M}_S(K^{*0}\bar{K^0})=\mathcal{M}_S(\bar{K^{*0}}K^0)
=g_{J/\psi(\psi^\prime)}\xi {\cal {F}}({\bf P}) \nonumber\\
\mathcal{M}_S(\omega\eta)&=&X_\eta
g_{J/\psi(\psi^\prime)}(1+2r){\cal
{F}}({\bf P})+Y_\eta\sqrt{2}\xi rg_{J/\psi(\psi^\prime)}{\cal {F}}({\bf P}) \nonumber\\
\mathcal{M}_S(\omega\eta^\prime)&=&X_{\eta^\prime}
g_{J/\psi(\psi^\prime)}(1+2r){\cal
{F}}({\bf P})+Y_{\eta^\prime}\sqrt{2}\xi rg_{J/\psi(\psi^\prime)}{\cal {F}}({\bf P}) \nonumber\\
\mathcal{M}_S(\phi\eta)&=& X_{\eta}\sqrt{2}\xi
rg_{J/\psi(\prime^\prime)}{\cal {F}}({\bf P})+Y_\eta
g_{J/\psi(\psi^\prime)}(1+r)\xi^2 {\cal
{F}}({\bf P}) \nonumber\\
\mathcal{M}_S(\phi\eta^\prime)&=& X_{\eta^\prime}\sqrt{2}\xi
rg_{J/\psi(\prime^\prime)}{\cal {F}}({\bf P})+Y_{\eta^\prime}
g_{J/\psi(\psi^\prime)}(1+r)\xi^2 {\cal {F}}({\bf P}) ,
\end{eqnarray}
where $X_\eta(X_{\eta^\prime})$ and $Y_\eta$($Y_{\eta^\prime}$) are
mixing amplitudes between $(u\bar{u}+d\bar{d})/\sqrt{2}$ and
$s\bar{s}$ components within the $\eta$ and $\eta^\prime$
wavefunctions:
\begin{eqnarray}
\eta &=& X_\eta |u\bar{u}
 +d\bar{d}\rangle/\sqrt{2}+ Y_\eta |s\bar
s\rangle,
\nonumber\\
\eta^\prime &=& X_{\eta^\prime} |u\bar{u}
 +d\bar{d}\rangle/\sqrt{2} + Y_{\eta^\prime} |s
\bar s\rangle . \label{eq_eta}
\end{eqnarray}
For the unitary $2\times 2$ mixing, we have
$X_\eta=Y_{\eta^\prime}=\cos\alpha_P$ and
$X_{\eta^\prime}=-Y_\eta=\sin\alpha_P$
 with $\alpha_P\equiv \theta_P +
\arctan(\sqrt{2})$.  The pseudoscalar mixing angle $\theta_P$ is in
a range of $-22^\circ\sim -13^\circ$.

For the decays of $J/\psi(\psi')\to \rho\pi$ and $K^*\bar{K}+c.c.$,
the short-distance amplitudes are rather simple as listed above. For
the decays into isoscalar final states, the situation would be
complicated by the DOZI process and glueball mixing. There have been
a lot of studies of the glueball mixing in the $\eta$ and $\eta'$
wavefunction~\cite{Li:2007ky,Escribano:2007cd,Cheng:2008ss,Tsai:2011dp},
which can contribute to the isoscalar decay channels.  However, in
this analysis we do not consider the glueball mixing effects since
the glueball components within $\eta$ and $\eta'$ are rather small
and need a delicate consideration. For the purpose of clarifying the
role played by the short-distance and long-distance transition
mechanisms in $J/\psi (\psi')\to VP$, we can leave the study of the
glueball mixing effects to be considered in a differently motivated
work~\cite{Tsai:2011dp}.

\subsection{Long-distance transition amplitudes via
IML}\label{sect-ii-3}

The IML transitions as a non-perturbative process seem to be a
natural mechanism to evade the OZI rule and HSR in the charmonium
decays~\cite{Liu:2009vv,Liu:2010um,Wang:2010iq,Zhang:2009kr}. The
relevant effective Lagrangians for the charmonium couplings to the
charmed mesons are as the
following~\cite{Colangelo:2003sa,Casalbuoni:1996pg}:
\begin{eqnarray}
\mathcal{L} = i \frac {g_2} {2} Tr[R_{c\bar{c}} \bar{H}_{2i}\gamma^\mu
{\stackrel{\leftrightarrow}{\partial}}_\mu \bar{H}_{1i}] + h.c.,
\end{eqnarray}
where the $S$-wave $J/\psi$ and $\psi'$ charmonium states are
expressed as
\begin{eqnarray}
R_{c\bar{c}}=  \left( \frac{1+ \rlap{/}{v} }{2} \right) (\psi^\mu
\gamma_\mu-\eta_c \gamma_5) \left( \frac{1- \rlap{/}{v} }{2}
\right).
\end{eqnarray}
The charmed and anti-charmed meson triplet read
\begin{eqnarray}
H_{1i}&=&\left( \frac{1+ \rlap{/}{v} }{2} \right)
[\mathcal{D}_i^{*\mu}
\gamma_\mu -\mathcal{D}_i\gamma_5], \\
H_{2i}&=& [\bar{\mathcal{D}}_i^{*\mu} \gamma_\mu
-\bar{\mathcal{D}}_i\gamma_5]\left( \frac{1- \rlap{/}{v} }{2}
\right),
\end{eqnarray}
where $\mathcal{D}$ and $\mathcal{D}^*$ denote the pseudoscalar and
vector charmed meson fields respectively, i.e.
$\mathcal{D}^{(*)}=(D^{0(*)},D^{+(*)},D_s^{+(*)})$.

Consequently, the Lagrangian for the $S$-wave $J/\psi$ and $\psi'$
is
\begin{eqnarray}
\mathcal{L}_\psi &=& ig_{\psi \mathcal{D}^* \mathcal{D}^*} (g_{\mu\sigma}g_{\nu\rho} -g_{\mu\rho}g_{\nu\sigma}
+g_{\mu\nu}g_{\rho\sigma}) \psi^\mu D^{*\nu} {\overleftrightarrow \partial}^\rho \mathcal{D}^{*\sigma\dagger}\nonumber\\
&&- ig_{\psi \mathcal{D}\mathcal{D}}\psi_\mu \mathcal{D}
\overleftrightarrow {\partial}^\mu {\mathcal{D}}^{\dagger} - g_{\psi
\mathcal{D}^*\mathcal{D}}\varepsilon^{\mu\nu\alpha\beta}\partial_\mu
\psi_\nu(\mathcal{D}^*_\alpha\overleftrightarrow{\partial}_\beta
{\mathcal{D}}^{\dagger}
+\mathcal{D}\overleftrightarrow{\partial}_\alpha{\mathcal{D}}^{*\dagger}_\beta)
,
\end{eqnarray}

The Lagrangians relevant to the light vector and pseudoscalar mesons
are,
 \begin{eqnarray}
 {\cal L} &=& - ig_{\mathcal{D}^*\mathcal{D}\mathcal{P}}(\mathcal{D}^i\partial^\mu \mathcal{P}_{ij}
 \mathcal{D}_\mu^{*j\dagger}-\mathcal{D}_\mu^{*i}\partial^\mu \mathcal{P}_{ij}\mathcal{D}^{j\dagger})
 +{1\over 2}g_{\mathcal{D}^*\mathcal{D}^*\mathcal{P}}
 \varepsilon_{\mu\nu\alpha\beta}\,\mathcal{D}_i^{*\mu}\partial^\nu \mathcal{P}^{ij}
 {\overleftrightarrow \partial}{}^{\!\alpha} \mathcal{D}^{*\beta\dagger}_j \nonumber \\
 &-& ig_{\mathcal{DDV}} \mathcal{D}_i^\dagger {\overleftrightarrow \partial}_{\!\mu} \mathcal{D}^j(\mathcal{V}^\mu)^i_j
 -2f_{\mathcal{\mathcal{D^*DV}}} \epsilon_{\mu\nu\alpha\beta}
 (\partial^\mu V^\nu)^i_j
 (\mathcal{D}_i^\dagger{\overleftrightarrow \partial}{}^{\!\alpha} \mathcal{D}^{*\beta j}-\mathcal{D}_i^{*\beta\dagger}{\overleftrightarrow \partial}{}{\!^\alpha} \mathcal{D}^j)
 \nonumber\\
 &+& ig_{\mathcal{D^*D^*V}} \mathcal{D}^{*\nu\dagger}_i {\overleftrightarrow \partial}_{\!\mu} \mathcal{D}^{*j}_\nu(\mathcal{V}^\mu)^i_j
 +4if_{\mathcal{D^*D^*V}} \mathcal{D}^{*\dagger}_{i\mu}(\partial^\mu \mathcal{V}^\nu-\partial^\nu
\mathcal{ V}^\mu)^i_j \mathcal{D}^{*j}_\nu,
 \label{eq:LDDV}
 \end{eqnarray}
with the convention $\varepsilon^{0123}=1$, where $\mathcal{P}$ and
$\mathcal{\mathcal{V}}_\mu$ are $3\times 3$ matrices for the octet
pseudoscalar and nonet vector mesons, respectively,
\begin{eqnarray} \nonumber \mathcal{P}=\left(
\begin{array} {ccc}
\frac {\pi^0} {\sqrt {2}} + \frac {\eta} {\sqrt {6}} & \pi^+ & K^+ \\
\pi^- & -\frac {\pi^0} {\sqrt {2}} + \frac {\eta} {\sqrt {6}} & K^0 \\
K^-& {\bar K}^0 & -\sqrt {2\over 3} \eta \\
\end{array}\right),
\mathcal{V}=\left(\begin{array}{ccc}\frac{\rho^0} {\sqrt {2}}+\frac {\omega} {\sqrt {6}}&\rho^+ & K^{*+} \\
\rho^- & -\frac {\rho^0} {\sqrt {2}} + \frac {\omega} {\sqrt {2}} & K^{*0} \\
K^{*-}& {\bar K}^{*0} & \phi \\
\end{array}\right).
\end{eqnarray}

\begin{figure}
\begin{center}
\begin{tabular}{ccc}
\hspace{-4cm}
\includegraphics[scale=1.0]{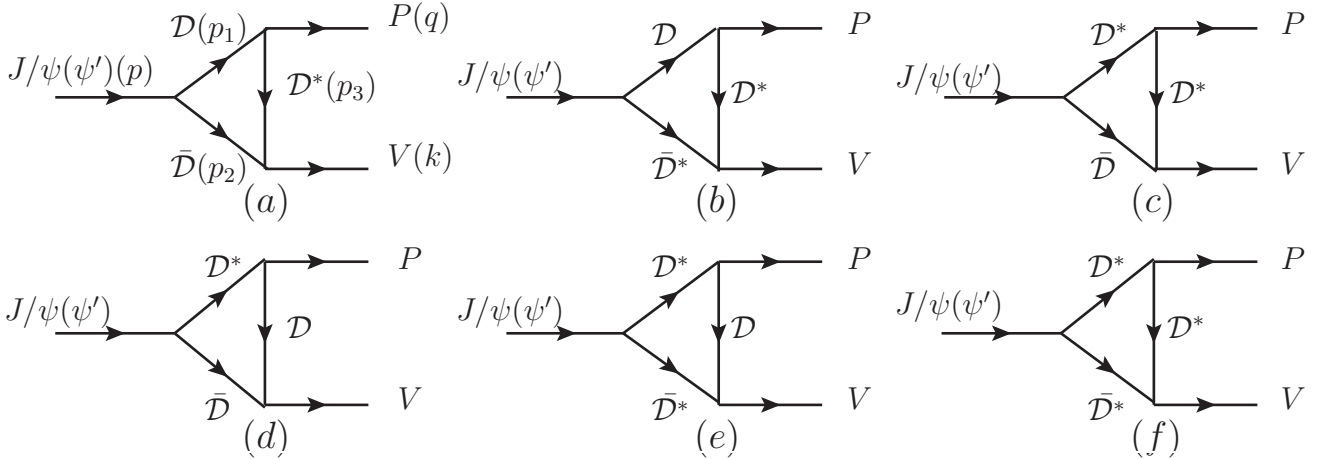}
%\put(-245,-10){(1)} \put(-125,-10){(2)} \put(-5,-10){(3)}
\vspace{-20cm}
\end{tabular}
\caption{Schematic diagrams for the long-distance IML transitions in
$J/\psi \ (\psi')\to VP$. In this case, the $c$ and $\bar{c}$
annihilate by multi-soft-gluon radiations and can be described by
intermediate charmed meson exchanges. The notations are
$\mathcal{\mathcal{D}}=(D^0, D^+,D_s^+)$ and
$\bar{\mathcal{D}}=(\bar{D}^0, D^-,D_s^-)$. The light flavors of the
charmed mesons are implicated by the final-state light mesons as
shown in Eq.~(\ref{eq-long}). }
 \label{long}\protect
\end{center}
\end{figure}

Based on the above Lagrangians, the explicit amplitudes in
Fig.~\ref{long} can be obtained
\begin{eqnarray}
\nonumber M_{\mathcal{D\bar{D}D^*}}&=&-4g_{\psi
\mathcal{DD}}g_{\mathcal{D^*DP}}f_{\mathcal{D^*DV}}\epsilon_{\psi}\cdot(p_2-p_1)\epsilon_{\mu\nu\alpha\beta}p_2^\mu\epsilon^\nu
p_3^\alpha q^\beta\\\nonumber
 M_{\mathcal{D\bar{D}^*D^*}}&=&-g_{\psi
\mathcal{DD^*}}g_{\mathcal{D^*DP}}\epsilon_{\mu\nu\rho\sigma}\epsilon_\psi^\mu
p_1^\rho p_2^\sigma q^\lambda
(-g_{\lambda\delta}+\frac{p_{3\lambda}p_{3\delta}}{m_{\mathcal{D}^*}^2})\\\nonumber
&\times
&(g_{\mathcal{D^*D^*V}}g^{\nu\delta}(p_2-p_3)\cdot\epsilon-4f_{\mathcal{D^*D^*V}}k^\delta\epsilon^\nu)\\\nonumber
M_{\mathcal{D^*\bar{D}D^*}}&=&-4g_{\psi
\mathcal{D^*D}}g_{\mathcal{D^*D^*P}}f_{\mathcal{D^*D^*V}}\epsilon_{\mu\rho\sigma\alpha}\epsilon^{\sigma
\lambda\kappa\tau}\epsilon_{abc\tau}\epsilon_\psi^\mu p_1^\rho
p_2^\alpha p_{1\lambda}p_{3\kappa} p_2^a \epsilon^bp_3^c\\\nonumber
M_{\mathcal{D^*\bar{D}D}}&=&-g_{\psi
\mathcal{D^*D}}g_{\mathcal{D^*DP}}g_{\mathcal{DDV}}\epsilon_{\mu\rho\sigma\alpha}\epsilon_\psi^\mu
p_1^\rho p_2^\alpha q^\sigma (p_2-p_3)\cdot \epsilon\\\nonumber
M_{\mathcal{D^*\bar{D}^*D}}&=&-g_{\psi \mathcal{D^*D^*}}
g_{\mathcal{D^*DP}}f_{\mathcal{D^*DV}}\epsilon\epsilon_{\mu\nu\alpha\beta}p_3^\mu
\epsilon^\nu p_2^\alpha q^\lambda
(-g_{\lambda\delta}+\frac{p_{2\lambda}p_{2\delta}}{m_{\mathcal{D}^*}^2})\\\nonumber
&\times &(\epsilon_\psi^\delta(p_1-p_2)^\beta-(p_1-p_2)\cdot
\epsilon
g^{\delta\beta}+\epsilon_\psi^\beta(p_1-p_2)^\delta)\\\nonumber
M_{\mathcal{D^*\bar{D}^*D^*}}&=&g_{\psi
\mathcal{D^*D^*}}g_{\mathcal{D^*D^*P}}\epsilon_{\mu\nu\alpha\alpha\beta}p_1^\nu
p_3^\alpha
(g_{\mathcal{\mathcal{D^*D^*V}}}g^{\beta\lambda}(p_2-p_3)\cdot\epsilon+4f_{\mathcal{D^*D^*V}}k^\beta\epsilon^\lambda)\\
&\times
&(-g_{\lambda\delta}+\frac{p_{2\lambda}p_{2\delta}}{m_{\mathcal{D}^*}^2})
(\epsilon_\psi^\mu(p_1-p_2)^\delta-\epsilon\cdot(p_1-p_2)g^{\mu\delta}+\epsilon\epsilon_\psi^\delta(p_1-p_2)^\mu),
\label{eq-amplitude}
\end{eqnarray}
where $p$, $k$, $q$ are the four-vector momenta of the incoming
charmonium, outgoing light vector, outgoing pseudoscalar,
respectively, and $p_1$, $p_2$, $p_3$ are the four-vector momenta of
the intermediate charmed mesons as denoted in Fig.~\ref{long}(a).
The subscriptions in the amplitudes denote the intermediate charmed
mesons in the loops, and we have omitted the denominators, form
factors, and integral measurement $\int\frac{d^4p_3}{(2\pi)^4}$ to
keep the formulaes short. The following couplings are adopted in the
numerical
calculations~\cite{Liu:2009vv,Liu:2010um,Zhang:2009kr,Wang:2010iq}:
\begin{eqnarray}
\label{eq:hsr}
g_{\psi\mathcal{DD}}=2g_2\sqrt{m_\psi}m_\mathcal{D},~~~g_{\psi
\mathcal{DD}^*}=\frac{g_{\psi
\mathcal{DD}}}{\widetilde{M}_\mathcal{D}},~~~ g_{\psi \mathcal{D}^*
\mathcal{D}^*}=g_{\psi
\mathcal{DD}^*}\sqrt{\frac{m_{\mathcal{D}^*}}{m_\mathcal{D}}}m_{\mathcal{D}^*},~~~
\widetilde{M}_\mathcal{D}=\sqrt{m_\mathcal{D}m_{\mathcal{D}^*}},
\end{eqnarray}
where $g_2=\frac{\sqrt{m_\psi}}{2m_\mathcal{D}f_\psi}$, and $m_\psi$
and $f_\psi=405$ MeV are the mass and decay constant of $J/\psi$.
The relative coupling strength of $\psi^\prime$ to $J/\psi$, i.e.
$g_{\psi^\prime \mathcal{D\bar{D}}}/g_{J/\psi
\mathcal{D\bar{D}}}=1$, is included as a input. The light meson
couplings to the charmed mesons are~\cite{Cheng:2004ru}
\begin{eqnarray}
\nonumber
g_{\mathcal{D^*DP}}=\frac{2g}{f_\pi}\sqrt{m_\mathcal{D}m_{\mathcal{D}^*}},~~~g_{\mathcal{D^*D^*P}}=\frac{g_{\mathcal{D^*DP}}}{\sqrt{m_\mathcal{D}m_{\mathcal{D}}^*}},\\
g_{\mathcal{DDV}}=g_{\mathcal{D^*D^*V}}=\frac{\beta_0
g_\mathcal{V}}{\sqrt{2}},~~~f_{\mathcal{D^*DV}}=\frac{f_{\mathcal{D^*D^*V}}}{m_{\mathcal{D}^*}}=\frac{\lambda
g_\mathcal{V}}{\sqrt{2}},~~~g_\mathcal{V}=\frac{m_\rho}{f_\pi},
\end{eqnarray}
where $g=0.59$, $\beta_0=0.9$, $\lambda=0.56$ GeV$^{-1}$ and
$f_\pi=132$ MeV are adopted.

The explicit amplitudes with different quantum number exchanges in
the loops have been given in Eq.~(\ref{eq-amplitude}). For each
decay mode the amplitude is dependent on the flavor component of the
final state light mesons. Thus, it is convenient to express the
flavor-dependent amplitudes as
\begin{eqnarray}
\nonumber
\mathcal{M}_{\rho\eta}&=&X_\eta\frac{1}{2}([\mathcal{D}^0\bar{\mathcal{D}}^0\mathcal{D}^0]
-[\mathcal{D}^+\mathcal{D}^+\mathcal{D}^-])+c.c.\\\nonumber
\mathcal{M}_{\rho\eta^\prime}&=&X_{\eta^\prime}\frac{1}{2}([\mathcal{D}^0\bar{\mathcal{D}}^0\mathcal{D}^0]
-[\mathcal{D}^+\mathcal{D}^+\mathcal{D}^-])+c.c.\\\nonumber
\mathcal{M}_{\omega\pi^0}&=&\frac{1}{2}([\mathcal{D}^0\bar{\mathcal{D}}^0\mathcal{D}^0]
-[\mathcal{D}^+\mathcal{D}^+\mathcal{D}^-])+c.c.\\\nonumber
\mathcal{M}_{\phi\pi^0}&=&0\\\nonumber
\mathcal{M}_{\rho^0\pi^0}&=&\frac{1}{2}([\mathcal{D}^0\bar{\mathcal{D}}^0\mathcal{D}^0]
+[\mathcal{D}^+\mathcal{D}^+\mathcal{D}^-]])+c.c.\\\nonumber
\mathcal{M}_{\rho^+\pi^-}&=&[\mathcal{D}^0\bar{\mathcal{D}}^0\mathcal{D}^++\mathcal{D}^-\mathcal{D}^+\bar{\mathcal{D}}^0]\\\nonumber
\mathcal{M}_{\omega\eta}&=&X_\eta\frac{1}{2}([\mathcal{D}^0\bar{\mathcal{D}}^0\mathcal{D}^0]
+[\mathcal{D}^+\mathcal{D}^+\mathcal{D}^-])+c.c.\\\nonumber
\mathcal{M}_{\omega\eta^\prime}&=&X_{\eta^\prime}\frac{1}{2}([\mathcal{D}^0\bar{\mathcal{D}}^0\mathcal{D}^0]
+[\mathcal{D}^+\mathcal{D}^+\mathcal{D}^-])+c.c.\\\nonumber
\mathcal{M}_{\phi\eta}&=&Y_\eta[\mathcal{D}_s^+\mathcal{D}_s^-\mathcal{D}_s^+]+c.c.\\\nonumber
\mathcal{M}_{\phi\eta^\prime}&=&Y_{\eta^\prime}[\mathcal{D}_s^+\mathcal{D}_s^-\mathcal{D}_s^+]+c.c.\\\nonumber
\mathcal{M}_{K^{*+}K^-}&=&[\mathcal{D}_s^-\mathcal{D}_s^+\bar{\mathcal{D}}^0]+[\mathcal{D}^0\bar{\mathcal{D}}^0\mathcal{D}_s^+]\\
\mathcal{\mathcal{M}}_{K^{*0}\bar{K}^0}&=&[\mathcal{D}_s^-\mathcal{D}_s^+\mathcal{D}^-]+[\mathcal{D}^+\mathcal{D}^-\mathcal{D}_s^+],
\label{eq-long}
\end{eqnarray}
where  $X_\eta$ ($X_{\eta^\prime}$) and $Y_\eta$ ($Y_{\eta^\prime}$)
have been defined earlier, and the amplitudes of $\rho^-\pi^+$,
$K^{*-}K^+$ and $\bar{K}^{*0}K^0$ have been implicated by their
conjugation channels listed above. Note that the destructive sign
between the charged and neutral meson loop amplitudes in those
isospin-violating channels, such as $\rho\eta$, $\rho\eta'$,
$\omega\pi^0$. The IML amplitudes for the $\phi\pi^0$ channel vanish
in the SU(3) symmetry limit.

Since the IML integrals are ultra-violet divergent, an empirical
tri-monopole form factor is introduced
\begin{eqnarray}
\mathcal{F}=\prod_i\frac{\Lambda_i^2-m_i^2}{\Lambda_i^2-p_i^2} \ ,
\end{eqnarray}
where $m_i$ is the mass of the exchanged particles and $p_i$ is the
corresponding four-vector momentum. As usual, $\Lambda_i$ is
parameterized into $\Lambda_i=m_i+\alpha\Lambda_{QCD}$ with
$\Lambda_{QCD}=0.22 \ \mbox{GeV}$ denoting the typical low energy
scale of QCD.

\section{Model results}

\subsection{Analyzing scheme}
\begin{figure}
\includegraphics[scale=0.6]{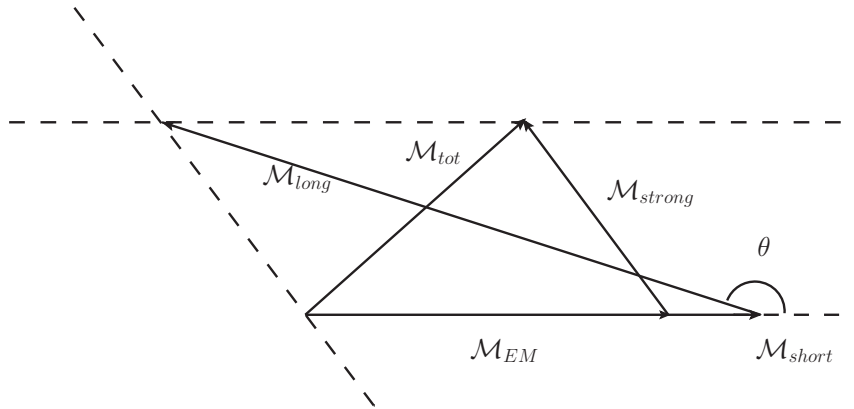}
\hspace{2cm}
%\put(-245,-10){(1)} \put(-125,-10){(2)} \put(-5,-10){(3)}
\vspace{-10cm} \caption{Decomposition of the transition amplitudes
for $J/\psi \ (\psi')\to VP$ in a complex plane. The EM amplitude is
assigned to point to the real axis while the short-distance
amplitude carries a trivial sign difference to the EM one. The
long-distance IML amplitude carries a phase due to hadronic effects.
The final summed amplitude $\mathcal{M}_{tot}$ is to be compared
with the experimental data. }
 \label{fit-scheme}
\end{figure}

As mentioned earlier, all the underlying mechanisms in the $VVP$
transitions would just contribute to the effective coupling
constant. This feature, on the one hand, can provide advantages for
disentangling different mechanisms, but on the other hand, may bring
difficulties to the numerical fittings since the final results would
only depend on the modulus of the summed amplitudes. Fortunately,
the dynamic features of those different transition mechanisms as
described in the previous section are useful for working out the
parameter fitting scheme and disentangling the underlying mechanisms
step by step. In Fig.~\ref{fit-scheme}, we illustrate the relation
between the EM and strong transition amplitudes (including the short
and long-distance ones) by the addition of vectors in the complex
plane. Our strategy of determining the amplitudes of those three
transition processes is as follows:

i) We treat the EM amplitude of each decay channel as a fixed vector
in the complex plane pointing to the real axis as shown by
Fig.~\ref{fit-scheme}.

The EM amplitudes can be independently fixed by the data for those
isospin-violating channels, i.e. $J/\psi(\psi')\to \rho\eta$,
$\rho\eta'$, $\omega\pi^0$ and $\phi\pi^0$. The same parameter
$\Lambda_{EM}=0.542$ GeV are then adopted for other decay channels
as a reliable estimate of the EM amplitudes in the VMD
model~\cite{Li:2007ky}.

It should be noted that as discussed in Ref.~\cite{Li:2007ky}, the
branching ratio fractions between $\psi'$ and $J/\psi$ decays into
these isospin-violating channels are approximately within the range
of 12\% rule. This is an indication that for a
single-mechanism-dominant process, the branching ratio fractions
still serves as a probe for the wavefunction at the origin. In
another word, if other mechanisms play a role, interferences among
those processes would break down the pQCD relation. Such a scenario
would be a natural explanation for the deviations observed in other
channels. For instance, in the $\rho\pi$ channel the interferences
between the EM and strong amplitudes would lead to significant
deviations from the 12\% rule. Our focus in this work is to
understand why the strong amplitude becomes compatible with the EM
one in such an HSR-violating channel.

ii) For the strong amplitudes including the short-distance and
long-distance IML amplitudes, it is reasonable to impose that the EM
and short-distance amplitude have a trivial relative phase since
both probe the charmonium wavefunctions at the origin and both are
real numbers. Moreover, for the short-distance SOZI amplitudes, we
impose a constraint to require that their exclusive contributions
should respect the 12\% rule, i.e. the magnitude of the
short-distance amplitudes can be treated as an input.
Equation~(\ref{total-trans}) can be rewritten as
\begin{eqnarray}\label{total-trans-2}
\mathcal{M}_{tot}\equiv
\mathcal{M}_{EM}+e^{i\delta}\mathcal{M}_{strong} ,
\end{eqnarray}
where $\mathcal{M}_{strong}$ is the amplitude for the total strong
transitions with a relative phase angle $\delta$ relative to the EM
one. By an overall fit of the experimental data~\cite{Li:2007ky},
the values of $\mathcal{M}_{strong}$ and $\delta$ can be fixed for
each channel. Then, with the fixed magnitude and direction for the
EM and short-distance amplitudes, the decomposition of
Eq.~(\ref{total-trans-2}) will allow us to determine the magnitude
and direction of the long-distance amplitude as shown by
Fig.~\ref{fit-scheme}.

We note that the overall fit of phase angle $\delta$ in
Ref.~\cite{Li:2007ky} suggests that  all the $VP$ channels in
$J/\psi$ or $\psi'$ decays share the same value of $\delta$.
Consequence of such an implication is a constraint on the magnitude
and direction for the long-distance amplitudes in each $VP$ channel.
What we are going to examine in the following part is the range of
the form factor parameter $\alpha$ in the meson loops, namely,
whether all the $VP$ channels share the same value of $\alpha$ at
all. Confirmation of such a scenario should be evidence for the
important contributions from the IML in $J/\psi$ and $\psi'$ decays.

Following the above procedure, we first consider $\rho\pi$ and
$K^*\bar{K}+c.c.$ decay channels. Since the final state light mesons
in these channels carry non-zero isospin, the short-distance
transitions can only occur via the SOZI process while the DOZI
process is forbidden. The analysis of these channels will then be
able to expose the interfering feature of the IML.

\subsection{Parameters and results}

The parameters to appear in the analysis include the followings: i)
the universal EM cut-off energy $\Lambda_{EM}=0.542$ GeV determined
by the isospin-violating channels. ii) the short-distance transition
strength $g_{J/\psi}=1.75\times 10^{-2}$ in the $J/\psi$ decay. This
is an input for the short-distance amplitudes. It determines the
short-distance transition strength $g_{\psi'}=1.25\times 10^{-2}$.
Therefore, the exclusive contributions from the short-distance
transitions still respect the $12\%$ rule. iii) the form factor
parameters $\alpha_{J/\psi}$ and $\alpha_{\psi^\prime}$ for the IML
transitions which determine the long-distance coupling strengths.
iv) the phase angles $\theta_{J/\psi}$ and $\theta_{\psi^\prime}$
between the EM and strong amplitudes  in the $J/\psi$ and
$\psi^\prime$ decays, respectively, as defined in
Eq.~(\ref{total-trans-2}).

Other implicated parameters such as the SU(3) flavor symmetry
breaking parameter $\xi = f_\pi/f_K\simeq 0.838$, vertex coupling
constants for the IML in Sec.~\ref{sect-ii-3}, and flavor mixing
angle $\theta_P=-22^\circ$ for $\eta$ and $\eta'$, in principle,
have been determined by independent processes.

In Table~\ref{tab-parameter}, parameters adopted in the calculations
are listed. As described in the above, some of those are treated as
input, while the three parameters, i.e. the IML form factor
parameter $\alpha$, phase angle $\theta$ and SU(3) flavor symmetry
parameter $\xi$, are fitted by the experimental data for $J/\psi$
and $\psi'\to \rho\pi$ and $K^*\bar{K}+c.c.$, respectively. When the
isospin zero decay channels are included, such as $J/\psi$ and
$\psi'\to \omega\eta$, $\omega\eta'$ etc, the DOZI parameter $r$ can
also be fitted. We also list the $\chi^2$ for the $J/\psi$ and
$\psi'$ decays, respectively. Briefly speaking, with the small
number of parameters we can achieve a reasonable description of the
overall experimental data, and the parameter uncertainties have been
well constrained. The relatively large $\chi^2$ value for the
$J/\psi$ decays are due to the relatively small experimental errors
in this channel. One notices that the value of $\alpha_{J/\psi}$ has
relatively large uncertainties in comparison with that of the
$\alpha_{\psi'}$. This is because the IML contributions are
relatively small in the $J/\psi$ decays, and insensitive to the form
factor parameter.

\begin{table}
\caption{Parameters fitted respectively by experimental data for
$J/\psi$ and $\psi^\prime\to VP$ in our analysis scheme.}
\begin{tabular}{ccc}
  \hline\hline
  parameter & $J/\psi$ & $\psi^\prime$ \\\hline
%  $g_s$ & $1.75\times 10^{-2}$ & $1.25\times 10^{-2}$ \\\hline
  $\xi$ & $0.71\pm0.017$ & $0.92\pm0.053$ \\\hline
  $r$ & $-0.201\pm0.006$ & $-0.097\pm0.024$ \\\hline
  $\theta$ & $45.0^\circ\pm 7.0^\circ$ & $174.6^\circ\pm 2.9^\circ$ \\\hline
  $\alpha$ & $0.09\pm0.03$ & $0.35\pm0.01$ \\\hline
  $\chi^2$&$52.2$&$9.17$\\
  %\hline
  \hline\hline
\end{tabular}
\label{tab-parameter}
\end{table}

\subsubsection{Isospin nonzero channels}

In Table~\ref{tab-nonzero}, the effective couplings extracted from
different transition mechanisms in $J/\psi$ and $\psi'\to \rho\pi$
and $K^*\bar{K}+c.c.$ are listed. Comparing the coupling $g_{short}$
in these two channels, one notices that the SU(3) flavor symmetry
breaking is at the scale of $\xi =0.71\sim 0.92$ which is compatible
with $\xi= f_\pi/f_K\simeq 0.838$. It shows that a small
long-distance contribution from the IML will optimize the
description of the data. The fitted form factor parameter
$\alpha_{J/\psi}\simeq 0.09$  is adopted for all exclusive decay
channels, which suggests a universal role played by the IML in
$J/\psi\to VP$.  The small contribution from the IML is
understandable since the mass of $J/\psi$ is much below the open
charm threshold. Therefore, it does not experience the long-distance
IML effects in the transition.

In $J/\psi\to VP$, the relatively small EM amplitudes implies
insignificant interferences between the EM and strong transition
amplitudes. This feature is indicated by the relatively small charge
asymmetries between $J/\psi\to K^{*+}K^-+c.c.$ and $J/\psi\to
K^{*0}\bar{K}^0+c.c.$

For $\psi'\to VP$, the short-distance coupling strength is
determined by the $12\%$ rule relation. A proper description of the
data leads to the determination of the long-distance IML amplitudes
as listed in Table~\ref{tab-nonzero}. Since the mass of the $\psi'$
is much closer to the open charm threshold, the long-distance IML
amplitudes become sizeable and play an important role in $\psi'\to
VP$. Given that the IML amplitudes are compatible with the
short-distance ones in magnitude, the destructive interferences
between the short and long-distance strong amplitudes have thus
significantly lowered the strong transition amplitudes to be
compatible with the EM ones. As a consequence, the further
interferences with the EM amplitudes lead to the significant charge
asymmetries between the branching ratios of $\psi'\to
K^{*0}\bar{K^0}+c.c.$ and $\psi'\to K^{*+}K^-+c.c.$

\begin{table}
\caption{The effective couplings (in unit of GeV$^{-1}$) extracted
from different transition mechanisms in $J/\psi \ (\psi')\to
\rho\pi$ and $K^*\bar{K}+c.c.$ }
\begin{tabular}{cccccc}
  \hline\hline
  $J/\psi\to VP$ & $g_{EM}$ & $g_{short}$ & $g_{long}$ & $|g_{strong}|$ & $|g_{tot}|$
  \\\hline
  $\rho^0\pi^0$ & $-2.26\times 10^{-4}$ & $-2.36\times 10^{-3}$ & $-1.60\times 10^{-5}$ & $2.37\times 10^{-3}$ &
  $2.60 \times 10^{-3}$\\\hline
  $K^{*+}K^-$ & $-2.17\times 10^{-4}$ & $-1.78\times 10^{-3}$ & $-1.45 \times 10^{-5}$ & $1.79 \times 10^{-3}$ &
  $2.0 \times 10^{-3}$
  \\\hline
  $K^{*0}\bar{K}^0$ & $3.28\times 10^{-4}$ & $-1.78\times 10^{-3}$ & $-1.45\times 10^{-5}$ & $1.79 \times 10^{-3}$ &
  $1.46 \times 10^{-3}$
  \\\hline\hline
  $\psi^\prime \to VP$  \\
  \hline
  $\rho^0\pi^0$ & $-9.58 \times 10^{-5}$ & $-1.11\times 10^{-3}$ & $-1.29\times 10^{-3}$ & $2.11\times 10^{-4}$ &
  $1.43\times 10^{-4}$
  \\\hline
  $K^{*+}K^-$ & $-9.07\times 10^{-5}$ & $-1.07\times 10^{-3}$ & $-1.25\times 10^{-3}$ & $2.07 \times 10^{-4}$ &
  $1.41\times 10^{-4}$
  \\\hline
  $K^{*0}\bar{K}^0$ & $1.37\times 10^{-4}$ & $-1.07\times 10^{-3}$ & $-1.28\times 10^{-3}$ & $2.33 \times 10^{-4}$ & $3.58 \times 10^{-4}$ \\
  \hline\hline
\end{tabular}
\label{tab-nonzero}
\end{table}

\subsubsection{Isospin zero channels}

For the isospin zero decay channels, such as $J/\psi (\psi')\to
\omega\eta$, $\omega\eta'$, $\phi\eta$, and $\phi\eta'$, the DOZI
transitions may contribute and two additional parameters have to be
included~\cite{Li:2007ky}. One is the $\eta$-$\eta'$ mixing angle
$\alpha_P$ defined in Eq.~(\ref{eq_eta}) and the other is the DOZI
coupling strength $r$ defined in Eq.~(\ref{eq_dozi}). We adopt the
commonly used value $\alpha_P=32.7^\circ$ as an input, while treat
$r$ as a free parameter to be determined by the isospin zero decay
channels. Meanwhile, all the other parameters determined in $J/\psi
\ (\psi')\to \rho\pi$ and $K^*\bar{K}+c.c.$ are fixed.

Eventually, it cannot be regarded as an overall fitting, and we do
not expect a perfect description of the data for $J/\psi (\psi')\to
\omega\eta$, $\omega\eta'$, $\phi\eta$, and $\phi\eta'$. This is
mainly because the involvement of the DOZI mechanism and possible
glueball mixing in the isospin zero channels should be considered in
a more delicate way. Therefore, we only expect that those isospin
zero channels to be described to the correct order of magnitude.

In Table~\ref{tab-results}, the model calculations of the branching
ratios of $J/\psi$ and $\psi'\to VP$ are listed in comparison with
the experimental values. The exclusive contributions from the EM,
short-distance and long-distance IML transitions, and the combined
strong contributions are also shown. The following points can be
learned:

i) For the isospin-violating channels, i.e. $J/\psi \ (\psi')\to
\rho\eta$, $\rho\eta'$ etc, the charged and neutral meson loops
would cancel out exactly in the isospin symmetry limit. In other
words, because of the isospin symmetry breaking, the mass difference
between the $u$ and $d$ quark leads to $m_{\cal D}^{(*)\pm} \neq
m_{\cal D}^{(*)0}$. As a result, the charged and neutral meson loops
cannot completely cancel out, and the residue part will contribute
to the isospin-violating amplitudes. Interestingly, we find that
contributions from such a mechanism is much smaller than the EM
transitions, which makes the isospin-violating channels ideal for
the test of the 12\% rule. In Table~\ref{tab-ratio}, the branching
ratio fraction $R$ is listed for all the $VP$ channels. One can see
that the 12\% rule is reasonably respected in those
isospin-violating channels.

ii) For the final-state isospin nonzero channels, the systematic
feature is that the long-distance amplitudes in $J/\psi \to \rho\pi$
and $K^*\bar{K}+c.c.$ are negligibly small and the strong amplitudes
are dominated by the short-distance ones. In contrast, the strong
amplitudes are suppressed in $\psi' \to \rho\pi$ and
$K^*\bar{K}+c.c.$ because of the destructive cancellations between
the short and long-distance amplitudes and become compatible with
the EM ones. The observed branching ratio fractions are then further
deviated from the 12\% rule by the interferences between the EM and
suppressed strong transition amplitudes in the $\psi'$ decays. We
have shown that the suppression of the strong amplitudes is due to
the open charm threshold effects via the IML transitions. The
branching ratio fraction $R$ is also listed in Table~\ref{tab-ratio}
to compare with the data. Moreover, with the exclusive
short-distance transition satisfying the 12\% rule in the $\rho\pi$
channel as a condition, other exclusive short-distance contributions
also satisfy it fairly well.

iii) For the final-state isospin zero channels, i.e. $J/\psi \
(\psi')\to \omega\eta$ and $\omega\eta'$ etc, the experimental data
have relatively large uncertainties and can be accounted for
approximately to the same order of magnitude. Similar to the study
of Ref.~\cite{Li:2007ky}, the DOZI contributions are found
necessary. In this analysis we do not consider the glueball mixing
effects in the $\eta$ and $\eta'$ wavefunctions since even though
there might be glueball components within the $\eta$ and $\eta'$,
uncertainties caused by them may not be as large as other sources
such as the DOZI contributions. In these processes, the EM
amplitudes are relatively smaller than the strong ones. But the
interferences among those strong transition amplitudes turn out to
be sensitive. More delicate treatment for those isospin-zero
channels are needed in further studies. One notices that the
branching ratio fraction $R$ can be reasonably accounted for except
for the channels involving $\eta'$. This might be an indication that
additional mechanisms should be considered.

\begin{table}
\caption{Theoretical results for the branching ratios of $J/\psi \
(\psi')\to VP$ calculated in our model. The experimental data are
from PDG2010~\cite{Nakamura:2010zzi}. }
\begin{tabular}{ccccccc}
  \hline\hline
  $BR(J/\psi\to VP)$ & EM & short-distance & long-distance & strong & total & exp. \\
  \hline
  $\rho\eta$ & $1.81\times10^{-4}$ & $0$ & $2.34\times10^{-12}$ &$2.34\times 10^{-12}$& $1.81\times10^{-4}$ & $(1.93\pm0.23)\times10^{-4}$ \\
  $\rho\eta^\prime$ & $1.37\times10^{-4}$ & $0$ & $2.21\times10^{-12}$ &$2.21\times 10^{-12}$& $1.37\times10^{-4}$ & $(1.05\pm0.18)\times10^{-4}$ \\
  $\omega\pi^0$ & $3.1\times10^{-4}$ & $0$ & $2.38\times10^{-12}$&$2.38\times 10^{-12}$& $3.10\times10^{-4}$ & $(4.5\pm 0.5)\times10^{-4}$ \\
 $\phi\pi^0$ &$9.52\times10^{-7}$ & $0$ & $0$ &0& $9.52\times10^{-7}$ & $<6.4\times10^{-6}$ \\
  $\rho^0\pi^0$ & $4.44\times10^{-5}$ & $4.85\times10^{-3}$ & $2.24\times10^{-7}$ &$4.89\times 10^{-3}$& $5.87\times10^{-3}$ & $(5.6\pm0.7)\times10^{-3}$ \\
  $\rho\pi$ & $1.06\times10^{-4}$ & $1.45\times10^{-2}$ & $6.71\times10^{-7}$ &$1.47\times 10^{-2}$& $1.73\times10^{-2}$ & $(1.69\pm0.15)\times10^{-2}$ \\
  $K^{*+}K^-+c.c.$ & $6.97\times10^{-5}$ & $4.69\times10^{-3}$ & $3.14\times10^{-7}$ &$4.74\times 10^{-3}$& $5.96\times10^{-3}$ & $(5.12\pm0.3)\times10^{-3}$ \\
  $K^{*0}\bar{K}^0+c.c.$ & $1.59\times10^{-4}$ & $4.68\times10^{-3}$ & $3.11\times10^{-7}$ &$4.73\times 10^{-3}$& $3.16\times10^{-3}$ & $(4.39\pm0.31)\times10^{-3}$ \\
  $\omega\eta$ & $1.4\times10^{-5}$ & $1.76\times10^{-3}$ & $1.50\times10^{-7}$ &$1.78\times 10^{-3}$& $2.11\times10^{-3}$ & $(1.74\pm0.20)\times10^{-3}$ \\
  $\omega\eta^\prime$ & $1.4\times10^{-5}$ & $9.91\times10^{-5}$ & $5.42\times10^{-8}$ &$1.02\times 10^{-4}$& $1.92\times10^{-4}$ & $(1.82\pm0.21)\times10^{-4}$ \\
  $\phi\eta$ & $2.35\times10^{-5}$ & $6.70\times10^{-4}$ & $3.22\times10^{-8}$ &$6.76\times 10^{-4}$& $9.52\times10^{-4}$ & $(7.5\pm0.8)\times10^{-4}$ \\
  $\phi\eta^\prime$ & $2.10\times10^{-5}$ & $2.07\times10^{-4}$ & $6.45\times10^{-8}$ &$2.12\times10^{-4}$& $9.93\times10^{-5}$ & $(4.0\pm0.7)\times10^{-4}$ \\
  \hline\hline
  $BR(\psi^\prime\to VP)$ \\
  \hline
  $\rho\eta$ & $1.42\times10^{-5}$ & $0$ & $4.13\times10^{-7}$ &$4.13\times 10^{-7}$& $1.94\times10^{-5}$ & $(2.2\pm0.6)\times10^{-5}$ \\
  $\rho\eta^\prime$ & $1.04\times10^{-5}$ & $0$ & $3.89\times10^{-7}$ &$3.89\times 10^{-7}$& $1.48\times10^{-5}$ & $(1.9^{+1.7}_{-1.2})\times10^{-5}$ \\
  $\omega\pi^0$ & $2.98\times10^{-5}$  & $0$ & $4.25\times10^{-7}$ &$4.25\times 10^{-7}$& $3.73\times10^{-5}$ & $(2.1\pm0.6)\times10^{-5}$ \\
  $\phi\pi^0$ & $9.78\times10^{-8}$ & $0$ & $0$&0& $9.78\times10^{-8}$ & $<4.0\times10^{-6}$ \\
  $\rho^0\pi^0$ & $4.36\times10^{-6}$ & $5.81\times10^{-4}$ & $7.85\times10^{-4}$ &$2.12\times 10^{-5}$& $9.72\times10^{-6}$ & *** \\
  $\rho\pi$ & $1.02\times10^{-5}$ & $1.74\times10^{-3}$ & $2.36\times10^{-3}$ &$6.36\times 10^{-5}$& $3.20\times10^{-5}$ & $(3.2\pm1.2)\times10^{-5}$ \\
  $K^{*+}K^-+c.c.$ & $7.03\times10^{-6}$ &$9.81\times10^{-4}$ & $1.33\times10^{-3}$ &$3.64\times 10^{-5}$& $1.70\times10^{-5}$ & $(1.7^{+0.8}_{-0.7})\times10^{-5}$ \\
  $K^{*0}\bar{K}^0+c.c.$ & $1.61\times10^{-5}$ & $9.81\times10^{-4}$ & $1.39\times10^{-3}$ &$4.61\times 10^{-5}$& $1.09\times10^{-4}$ & $(1.09\pm0.20)\times10^{-4}$ \\
  $\omega\eta$ & $1.10\times10^{-6}$ & $3.24\times10^{-4}$ & $5.57\times10^{-4}$ &$3.52\times 10^{-5}$& $2.48\times10^{-5}$ & $<1.1\times10^{-5}$ \\
  $\omega\eta^\prime$ & $1.12\times10^{-6}$ & $6.23\times10^{-5}$ & $2.31\times10^{-4}$ &$5.43\times 10^{-5}$& $4.01\times10^{-5}$ & $(3.2^{+2.5}_{-2.1})\times10^{-5}$ \\
  $\phi\eta$ & $2.26\times10^{-6}$  & $1.55\times10^{-4}$ & $1.73\times10^{-4}$ &$1.92\times 10^{-6}$& $2.25\times10^{-6}$ & $(2.8^{+1.0}_{-0.8})\times10^{-5}$ \\
  $\phi\eta^\prime$ & $2.22\times10^{-6}$ & $1.85\times10^{-4}$ & $3.99\times10^{-4}$ &$4.33\times 10^{-5}$& $6.42\times10^{-5}$ & $(3.1\pm1.6)\times10^{-5}$ \\
  \hline\hline
\end{tabular}
\label{tab-results}
\end{table}

\begin{table}\footnotesize
\caption{The branching ratio fraction $R=BR(\psi^\prime\to
VP)/BR(J/\psi\to VP)$ given by our model (total) and the exclusive
short-distance transitions. For the isospin-violating channels, the
ratios are dominated by the EM transitions. The dash ``-" means the
transition is either absent or the data are not available. The
experimental ratios are listed as a comparison.} \hspace{0cm}
\begin{tabular}{c|c|c|c|c|c|c|c|c|c|c|c|c}
  \hline\hline
VP mode$(\%)$ & $\rho\eta$ & $\rho\eta^\prime$ & $\omega\pi^0$ &  $\phi\pi^0$ & $\rho^0\pi^0$ & $\rho\pi$ & $K^{*+}K^-+c.c.$ & $K^{*0}\bar{K}^0+c.c.$ & $\omega\eta$ & $\omega\eta^\prime$ & $\phi\eta$ &  $\phi\eta^\prime$  \\
  \hline
  short-distance & - & - & - & - & 11.98 & 12.0 & 20.92 &
20.96 & 18.41 & 62.87& 23.13 & 89.37
  \\\hline
  total & $10.72$ & $10.80$ & $12.03$ & $10.27$ & $0.17$ & $0.18$ & $0.29$  & $3.45$ & $1.18$ & $20.89$ & $0.24$ & $64.65$
  \\\hline
  \multirow{2}{*}{experiment}
 & $7.40$ & $5.69$ &$3.0$ & \multirow{2}{*}{-} & \multirow{2}{*}{-} & $0.11$ & $0.19$ & $1.89 $ & \multirow{2}{*}{$<0.63$} & $5.42$ & $2.41 $ & $3.19$ \\
  &$\sim 16.50$&$\sim 41.38$&$\sim 6.75$&&&$\sim 0.29$&$\sim 0.52$&$\sim
  3.16$&&$\sim 35.40$&$\sim 5.67$&$\sim 14.24$\\
  \hline\hline
\end{tabular}
\label{tab-ratio}
\end{table}

\section{Summary}

By systematically analyzing the transition mechanisms for $J/\psi$
and $\psi'\to VP$, we have shown that the long-distance IML
transitions are crucial for our understanding of the long-standing
``$\rho\pi$ puzzle". Since the mass of $\psi'$ is close to the open
charm threshold, its decays into $VP$ are affected significantly via
the IML transitions. In particular, the long-distance IML
transitions provide a mechanism to evade the pQCD HSR and their
destructive interferences with the short-distance amplitudes in the
$\psi'$ decays cause apparent deviations from the ``12\% rule". The
IML transition turns out to be a rather general nonperturbative
mechanism in the charmonium energy region. Our analysis suggests
that this mechanism should be present in all the decay modes. The
same coincident cancelation between the short and long-distance
amplitudes also causes large charge asymmetries between $\psi'\to
K^{*+}K^-+c.c.$ and $K^{*0}\bar{K}^0+c.c.$

It should be addressed that the open charm threshold effects via the
IMLs can also contribute to the process of $\psi(3770)\to VP$. As
shown in Refs.~\cite{Zhang:2009kr,Liu:2009dr}, the IML mechanism can
be a natural explanation for the sizeable $\psi(3770)$
non-$D\bar{D}$ decay branching ratios observed in
experiment~\cite{He:2005bs,:2007zt,Besson:2005hm,Ablikim:2006zq,Ablikim:2006aj}.

As a manifestation of the open charm threshold effects, the IML
mechanism may also play an important role in other decay modes, such
as $J/\psi (\psi')\to VS$, $VT$, and $PP$ etc. As the spin partners
of $J/\psi (\psi')$, the study of the ratio of $BR(\eta_c'\to
VV)/BR(\eta_c\to VV)$ should also be useful for clarifying the role
played by the IML and provide some insights into the long-standing
``$\rho\pi$ puzzle". Such a process has been investigated in
Ref.~\cite{Wang:2010iq} and recently updated in
Ref.~\cite{Wang:2012wj}. We expect that with the help of precise
measurements of various decay modes at BESIII, the IML mechanism can
be established as an important nonperturbative dynamics in the
charmonium energy region.

\acknowledgments

Authors thank Profs. S. Brodsky, K.-T. Chao, C.-H. Chang, T. Huang,
X.-Q. Li, and C.-Z. Yuan for useful discussions on this topic. This
work is supported, in part, by National Natural Science Foundation
of China (Grant No. 11035006), Chinese Academy of Sciences
(KJCX2-EW-N01), and Ministry of Science and Technology of China
(2009CB825200).

\end{document}